# Blind Estimation of Primary User Traffic Parameters Under Sensing Errors

Wesam Gabran*, Przemysław Pawełczak†, Chun-Hao Liu*, and Danijela Cabric*
*University of California, Los Angeles, Engineering IV Building, Los Angeles, CA 90095-1594, USA
†Fraunhofer Institute for Telecommunications, Heinrich Hertz Institute, Einsteinufer 37, 10587 Berlin, Germany
Email: {wgabran, liuch37, danijela}@ee.ucla.edu; przemyslaw.pawelczak@hhi.fraunhofer.de

*Abstract*—**In this work we investigate the bounds on the estimation accuracy of Primary User (PU) traffic parameters with exponentially distributed busy and idle times. We derive closed-form expressions for the Cramér-Rao bounds on the mean squared estimation error for the blind joint estimation of the PU traffic parameters, specifically, the duty cycle, and the mean arrival and departure rates. Moreover, we present the corresponding maximum-likelihood estimators for the traffic parameters. In addition, we derive a modified likelihood function for the joint estimation of traffic parameters when spectrum sensing errors are considered, and we present the impact of spectrum sensing errors on the estimation error via simulations. Finally, we consider a duty cycle estimator, common in traffic estimation literature, that is based on averaging the traffic samples. We derive, in closed-form, the mean squared estimation error of the considered estimator under spectrum sensing errors.**

## I. INTRODUCTION

In Dynamic Spectrum Access (DSA) networks Secondary Users (SUs) search for and utilize temporarily vacant licensed spectrum. The SUs have to sense for the presence of Primary Users (PUs) before accessing a spectral band using spectrum sensing [1]. The PU spectral utilization can be modeled as a stochastic process [2] where knowledge of the PU traffic statistics can improve the performance of SU channel selection algorithms and yield more efficient resource allocation. The benefit of estimating the PU traffic parameters is, however, limited by the accuracy of the estimates.

Recalling related work discussion from [3, Sec. I-B], the most notable work that analyzes the estimation accuracy of PU time-domain traffic parameters can be found in [3]–[7]. All considered works assume that PUs have exponentially distributed idle (off-) and busy (on-) times, for analytical tractability, where traffic is modeled in terms of the mean duty cycle, $u$, and the mean departure and arrival rates denoted by $\lambda_f$ and $\lambda_n$, respectively. In [4] traffic samples averaging was used for estimating $u$ while maximum-likelihood (ML) estimation was adopted for estimating $\lambda_f$. Meanwhile, the authors in [5] proposed using Bayesian estimation for estimating $\lambda_f$ and $\lambda_n$. However, both [4] and [5] did not quantify the estimation error. On the other hand, the authors in [6], [7] investigated a number of sampling schemes to minimize the estimation error when ML estimation is used for estimating $\lambda_f$. However, the authors assumed perfect knowledge of $u$, and did not provide a closed-form expression for the estimation error where the estimation error results were based on simulations. In [3], closed-form expressions for the Cramér-Rao (CR) bounds on the estimates of $u$ and $\lambda_f$ were derrived assuming perfect knowledge of $\lambda_f$ and $u$, respectively. Besides, expressions for the mean squared estimation error in $u$ when averaging is used under non-uniform sampling were derrived. Also the impact of spectrum sensing errors on the estimation error was taken into account. Unfortunately, in all aforementioned works [3]–[7], the fully-blind estimation error, where all traffic parameters are jointly estimated, was not considered. Instead, the authors assumed the availability of a priori knowledge regarding either $u$, $\lambda_f$, or $\lambda_n$ when estimating the traffic parameters, which in most practical cases is not a valid assumption. Furthermore, the effect of spectrum sensing errors on the estimation error was not considered for fully-blind traffic estimation.

To understand the importance of accurate PU traffic statistics estimation, the contribution of this work is twofold. Extending the work of [3], first, we derive closed-form expressions for the CR bounds on the mean squared estimation error for the joint estimation of $u$ and $\lambda_f$ when uniform sampling is employed. We also present the CR bounds for the joint estimation of $u$ and $\lambda_n$. Moreover, we show, via simulations, that the estimation error of the corresponding joint ML estimators achieve the CR bounds asymptotically. Second, we study the impact of spectrum sensing errors on the traffic parameters estimation accuracy. We present modified likelihood functions that put into account the spectrum sensing errors, and we present the effect of sensing errors on the performance of the joint ML estimators using simulations. Moreover, we derive a closed-form expression for the mean squared estimation error in $u$ when traffic sample averaging is used while considering spectrum sensing errors.

## II. SYSTEM MODEL

In this work, we follow the model introduced in [4], where we consider a single channel licensed to a single PU. The PU traffic, which is to be estimated, is assumed to be stationary over a sufficiently large time window with exponentially distributed off- and on-times. The probability density function

This work has been supported by the National Science Foundation under CNS grant 1117600 and the German Federal Ministry of Economics and Technology under grant 01ME11024.

of an exponentially distributed random variable, $x$, is given as $f_\lambda(x) = \lambda e^{-\lambda x}$, for $x \geq 0$ and $f_\lambda(x) = 0$, otherwise, where $\lambda$ denotes the rate parameter [8, Eq. (3.15)]. The distribution of the PU off- and on-times is denoted by $f_\lambda(x)$ with $\lambda = \lambda_f$ and $\lambda = \lambda_n$, respectively[1]. The mean PU off- and on-times are equal to the reciprocal of $\lambda_f$ and $\lambda_n$, respectively. Besides, the duty cycle $u$ of the PU can be calculated as [8, Sec. 11.3] $u = \frac{\lambda_f}{\lambda_f + \lambda_n}$. Hence, $\lambda_f$, $\lambda_n$ and $u$ are inter-dependent, where estimating any two of the three parameters is sufficient to completely estimate the PU traffic parameters.

The PU state is modeled as a semi-Markov process as in [4]. The model is used to formulate the PU state transition probabilities. Denote the PU state transition probability by $\Pr_{xy}(t)$, which corresponds to the probability that the PU state changes from state $x$ to state $y$ within time $t$, where $\{x, y\} = 0$ denotes that the PU is idle while $\{x, y\} = 1$ denotes that the PU is active. The PU state transition probabilities were derived in [4, Sec. 6.1] as

$$\Pr_{xy}(t) = \begin{cases} 1 - u + u e^{\frac{-\lambda_f t}{u}}, & x=0, y=0, \quad \text{(1a)} \\ 1 - \Pr_{00}(t), & x=0, y=1, \quad \text{(1b)} \\ u + (1-u) e^{\frac{-\lambda_f t}{u}}, & x=1, y=1, \quad \text{(1c)} \\ 1 - \Pr_{11}(t), & x=1, y=0. \quad \text{(1d)} \end{cases}$$

In this work $\Pr_{xy}(t)$ is later used to derive the mean squared error (MSE) in the estimated traffic parameters.

To estimate the traffic parameters, the channel is sampled in order to acquire data regarding the state of the PU (on- or off-). Denote the PU traffic samples by the vector $\mathbf{z} = [z_1, z_2, \cdots, z_N]$ where $N$ is the total number of samples, $z_n$ is the $n$th traffic sample, and $z_n = 1$ if the PU is active and $z_n = 0$, otherwise. Moreover, in the proposed model, we consider the general case where the spectrum sensing process is prone to errors. The sensing error is modeled in the form of false alarm and mis-detection probabilities, denoted by $P_f$ and $P_m$, respectively. The sensing error is assumed to be independent for the different traffic samples. The estimated PU traffic samples are denoted by the vector $\tilde{\mathbf{z}} = [\tilde{z}_1, \tilde{z}_2, \cdots, \tilde{z}_N]$ where $\tilde{z}_n$ is the $n$th estimated traffic sample. It follows that $\tilde{z}_n = 1$ if $z_n = 1$ and no mis-detection error occurred, or $z_n = 0$ and a false alarm error occurred. Similarly, $\tilde{z}_n = 0$ if $z_n = 1$ and a mis-detection error occurred, or $z_n = 0$ and no false alarm error occurred. Moreover, the inter-sample times are given by the vector $\mathcal{T} = [T_1, T_2, \cdots, T_{N-1}]$, where $T_n$ denotes the time between samples $z_n$ and $z_{n+1}$. Finally, the total observation window length is denoted by $T$, where $T = \sum_{n=1}^{N-1} T_n$.

Estimators of $u$, $\lambda_f$ and $\lambda_n$ were analytically described in closed-form in [3]–[7]. However, a measure of the estimation error when there is no a priori knowledge of $u$, $\lambda_f$ or $\lambda_n$ was not given. In the following section, we present the error bounds when blindly estimating the traffic parameters with no a priori information.

## III. CRAMÉR-RAO BOUNDS FOR THE JOINT ESTIMATION OF THE PRIMARY USER DUTY CYCLE $u$, DEPARTURE RATE $\lambda_f$, AND ARRIVAL RATE $\lambda_n$

As previously stated, the traffic parameters, $u$, $\lambda_f$, and $\lambda_n$ are inter-dependent, where estimating any two of the three parameters is sufficient to completely estimate all three parameters. We consider the case where $u$ and $\lambda_f$ are jointly estimated where we derive the CR bounds on the estimation error. The CR bound quantifies the minimum mean squared estimation error that can be achieved by any unbiased estimator. ML estimators achieve the CR lower bound as the sample size tends to infinity when certain regularity conditions are satisfied [14, Ch. 12]. Accordingly, we present expressions for the ML estimators when $u$ or $\lambda_f$ are estimated separately. Finally, we present the ML estimators and CR bounds when $u$ and $\lambda_n$ are jointly estimated.

For analytical tractability uniform sampling is assumed with a constant inter-sample time of $T_c$ seconds, where $T_c = T/(N-1)$. Besides, the sensing error probabilities, $P_f$ and $P_m$, are assumed to be equal to zero, hence, $\tilde{z}_n = z_n \forall n$. Note that the effect of sensing error on the estimation error is discussed in Section III-C. The likelihood function of the traffic samples given $u$ and $\lambda_f$ is derived in a similar manner to [4, Sec. 6.1] and can be written as

$$L(\mathbf{z}|u, \lambda_f) = \Pr(\mathbf{z}|u, \lambda_f) = \Pr(z_1|u, \lambda_f) \prod_{i=1}^{N-1} \Pr(z_{i+1}|z_i, u, \lambda_f)$$
$$= u^{z_1}(1-u)^{1-z_1} \prod_{i=1}^{N-1} \Pr_{z_i z_{i+1}}(T_c|u, \lambda_f), \quad (2)$$

where the Markovian property has been applied and $\Pr_{z_i z_{i+1}}(T_c|u, \lambda_f)$ denotes the probability that a sample $z_i$ is followed by a sample $z_{i+1}$ for the inter-sample time $T_c$, given $u$ and $\lambda_f$. Expression (2) can be written as

$$L(\mathbf{z}|u, \lambda_f) = u^{z_1}(1-u)^{1-z_1} \Pr_{00}^{n_0}(T_c|u, \lambda_f) \Pr_{01}^{n_1}(T_c|u, \lambda_f)$$
$$\times \Pr_{10}^{n_2}(T_c|u, \lambda_f) \Pr_{11}^{n_3}(T_c|u, \lambda_f), \quad (3)$$

where $n_0$, $n_1$, $n_2$ and $n_3$ denote the number of $(0 \to 0)$, $(0 \to 1)$, $(1 \to 0)$, and $(1 \to 1)$ PU state transitions, respectively, from the total of $N-1$ transitions among $N$ samples. Then, the ML estimators of $u$ and $\lambda_f$, denoted by $\tilde{u}_m$ and $\tilde{\lambda}_{f,m}$, respectively, can be found by simultaneously solving $\partial \log L(\mathbf{z}|\tilde{u}_m, \tilde{\lambda}_{f,m})/\partial \tilde{\lambda}_{f,m} = 0$ and $\partial \log L(\mathbf{z}|\tilde{u}_m, \tilde{\lambda}_{f,m})/\partial \tilde{u}_m = 0$, expressed respectively as

$$n_1 + n_2 = \frac{n_0 \Pr_{01}}{\Pr_{00}} + \frac{n_3 \Pr_{10}}{\Pr_{11}}, \quad (4)$$

$$\frac{z_1 - \tilde{u}_m}{1 - \tilde{u}_m} = \frac{[n_1 \Pr_{00} - n_0 \Pr_{01}]\left[\tilde{\Gamma}_{c,m} T_c \tilde{\lambda}_{f,m} - \Pr_{01}\right]}{\Pr_{00} \Pr_{01}}$$
$$+ \frac{[n_2 \Pr_{11} - n_3 \Pr_{10}]\left[\frac{1 - \tilde{u}_m}{\tilde{u}_m}\tilde{\Gamma}_{c,m} T_c \tilde{\lambda}_{f,m} + \Pr_{01}\right]}{\Pr_{10} \Pr_{11}}, \quad (5)$$

---
[1] Note that the assumption on the exponential distribution of off- and on-times is common in DSA literature, e.g., see recent examples of [9]–[11]; see also recent papers confirming the exponential distribution of time-domain utilization of certain licensed channels [2], [12], [13].

where $\tilde{\Gamma}_{c,m} = e^{\frac{-\tilde{\lambda}_{f,m} T_c}{\tilde{u}_m}}$, and the argument of $\Pr_{xy}(T_c|\tilde{u}_m, \tilde{\lambda}_{f,m})$, $x, y \in \{0,1\}$, in (4) and (5) is dropped for brevity. The estimators, $\tilde{u}_m$ and $\tilde{\lambda}_{f,m}$, cannot be written in a simple closed-form and are to be evaluated numerically.

### A. The CR Bound for Estimating $u$ and $\lambda_f$

The CR bounds for the estimators of $\boldsymbol{\theta} = [u, \lambda_f]^T$ are defined as the diagonal elements of the reciprocal of the 2-by-2 Fisher information matrix defined by

$$I_N(u, \lambda_f) = E_{\boldsymbol{z}}\left[\frac{-\partial^2}{\partial \boldsymbol{\theta}^2} \log L(\boldsymbol{z}|\boldsymbol{\theta})\right], \qquad (6)$$

where the subscript $N$ in $I_N$ implies that the Fisher information is calculated for a traffic sample vector of $N$ samples, and $E_{\boldsymbol{z}}[\cdot]$ is the expectation calculated over all permutations of $\boldsymbol{z}$. Denote the CR bounds for the estimators of $u$ and $\lambda_f$ by $V_{u,CR}$ and $V_{\lambda_f,CR}$, respectively. Hence, the bounds can be expressed as

$$V_{u,CR} = \frac{I_N(u, \lambda_f)[2,2]}{|I_N(u, \lambda_f)|}, \qquad (7)$$

and

$$V_{\lambda_f,CR} = \frac{I_N(u, \lambda_f)[1,1]}{|I_N(u, \lambda_f)|}. \qquad (8)$$

The elements of the Fisher information matrix can be written as follows (noting again that the argument of $\Pr_{xy}(T_c|u, \lambda_f)$, $x, y \in \{0,1\}$, is dropped for brevity)

$$I_N(u,\lambda_f)[1,1] = E_{\boldsymbol{z}}\left[\frac{-\partial^2}{\partial u^2}\log L(\boldsymbol{z}|\boldsymbol{\theta})\right] = \left(\Gamma_c^2 \lambda_f T_c(N-1)\right)$$
$$\times \frac{\lambda_f T_c(1-u)(1+\Gamma_c) + 2u(2u-1)(1-\Gamma_c)}{u^2 \Pr_{01}\Pr_{00}\Pr_{11}}$$
$$- \frac{(N-1)\Gamma_c^2 - N\Gamma_c + u\Pr_{10}((3N-2)\Gamma_c - N)}{u(1-u)\Pr_{00}\Pr_{11}}, \qquad (9)$$

$$I_N(u,\lambda_f)[1,2] = I_N(u,\lambda_f)[2,1] = E_{\boldsymbol{z}}\left[\frac{-\partial^2}{\partial u \partial \lambda_f}\log L(\boldsymbol{z}|\boldsymbol{\theta})\right]$$
$$= \frac{-(N-1)\Gamma_c^2 T_c[T_c\lambda_f(1-u)(1+\Gamma_c) + \Pr_{01}(2u-1)]}{u\Pr_{01}\Pr_{00}\Pr_{11}}, \qquad (10)$$

and

$$I_N(u,\lambda_f)[2,2] = E_{\boldsymbol{z}}\left[\frac{-\partial^2}{\partial \lambda_f^2}\log L(\boldsymbol{z}|\boldsymbol{\theta})\right]$$
$$= \frac{(N-1)\Gamma_c^2 T_c^2(1-u)(1+\Gamma_c)}{\Pr_{01}\Pr_{00}\Pr_{11}}, \qquad (11)$$

where $\Gamma_c = e^{\frac{-\lambda_f T_c}{u}}$. Expressions (9), (10), and (11) are derived in Appendix A, Appendix B, and Appendix C, respectively. The determinant of $I_N(u, \lambda_f)$ can be expressed as

$$|I_N(u, \lambda_f)| = \frac{(\Gamma_c T_c)^2(N-1)(2\Gamma_c + N(1-\Gamma_c))}{u\Pr_{01}\Pr_{00}\Pr_{11}}. \qquad (12)$$

Hence, using (9)–(12), the CR bounds for the estimators of $u$ and $\lambda_f$ can be expressed, respectively, as

$$V_{u,CR} = \frac{u(1-u)(1+\Gamma_c)}{N(1-\Gamma_c) + 2\Gamma_c}, \qquad (13)$$

and

$$V_{\lambda_f,CR} = \frac{\lambda_f[\lambda_f T_c(1-u)(1+\Gamma_c) + 2u(2u-1)(1-\Gamma_c)]}{uT_c(2\Gamma_c + N(1-\Gamma_c))}$$
$$- \frac{\Pr_{01}[u\Pr_{10}((3N-2)\Gamma_c - N) + (N-1)\Gamma_c^2 - \Gamma_c N]}{(\Gamma_c T_c)^2(1-u)(N-1)(2\Gamma_c + N(1-\Gamma_c))}. \qquad (14)$$

For a fixed observation window $T$, as $N$ increases, the CR bounds for the estimators of $u$ and $\lambda_f$ approach asymptotic values denoted by $V_{u,CR,L}$ and $V_{\lambda_f,CR,L}$, respectively. The lower bounds on the CR bounds can be derived as

$$V_{u,CR,L} = \lim_{N \to \infty} V_{u,CR} = \frac{u(1-u)}{1 + \frac{T\lambda_f}{2u}}, \qquad (15)$$

and

$$V_{\lambda_f,CR,L} = \lim_{N \to \infty} V_{\lambda_f,CR} = \frac{\lambda_f(u + T\lambda_f)}{T(1-u)(2u + T\lambda_f)}. \qquad (16)$$

Note that lower bound on the CR bounds can only be decreased by increasing the total observation window length $T$. Moreover, it has been shown in [3, Eq. (24)] that the asymptotic CR bound on the estimator of $u$ assuming perfect knowledge of $\lambda_f$ equals $u(1-u)/(1 + \lambda_f T/u)$. Comparing (15) with [3, Eq. (24)], it is clear that the expressions would match if the observation window length $T$ in (15) is multiplied by a factor of two. That is, the effect of having a priori knowledge regarding $\lambda_f$ on the asymptotic CR bound on estimating $u$ is equivalent to doubling the observation window length.

### B. The CR Bound for Estimating $u$ and $\lambda_n$

For the joint estimation of $u$ and $\lambda_n$, the CR bound on the estimation error of $u$ can also be expressed as (15) while the CR bound for the estimation error in $\lambda_n$ can be expressed as

$$V_{\lambda_n,CR} = \frac{\lambda_n[\lambda_n T_c u(1+\Gamma_c) + 2(1-u)(1-2u)(1-\Gamma_c)]}{(1-u)T_c(2\Gamma_c + N(1-\Gamma_c))}$$
$$- \frac{\Pr_{10}[u\Pr_{10}((3N-2)\Gamma_c - N) + (N-1)\Gamma_c^2 - \Gamma_c N]}{(\Gamma_c T_c)^2 u(N-1)(2\Gamma_c + N(1-\Gamma_c))}, \qquad (17)$$

where the derivations are omitted for brevity and follow that for the joint estimation of $u$ and $\lambda_f$. Finally, as $N$ increases for a fixed observation window $T$, $V_{\lambda_n,CR}$ approaches an asymptote, $V_{\lambda_n,CR,L}$, that can be derived as

$$V_{\lambda_n,CR,L} = \lim_{N \to \infty} V_{\lambda_n,CR} = \frac{\lambda_n((1-u) + T\lambda_n)}{Tu(2(1-u) + T\lambda_n)}. \qquad (18)$$

### C. The Impact of Spectrum Sensing Errors on Estimating $u$, $\lambda_f$, and $\lambda_n$

In the presence of spectrum sensing errors, any PU traffic samples vector $\boldsymbol{z}$ can result in an estimated PU traffic samples vector $\tilde{\boldsymbol{z}}$ with a non-zero probability. Hence, the likelihood function presented in (3) has to be modified in order to account for spectrum sensing errors. Define $\mathcal{Z} = [\mathcal{Z}_1, \mathcal{Z}_2, \cdots, \mathcal{Z}_{2^N}]$ as a vector containing all $2^N$ permutations of $\boldsymbol{z}$. Then, the modified likelihood function can be expressed as

$$L(\tilde{\boldsymbol{z}}|u, \lambda_f) = \sum_{n=1}^{2^N} \Pr(\mathcal{Z}_n|u, \lambda_f) S(\tilde{\boldsymbol{z}}|\mathcal{Z}_n), \qquad (19)$$

where $\Pr(\mathcal{Z}_n|u,\lambda_f)$ is the probability of occurrence of the PU traffic samples vector $\mathcal{Z}_n$ and equals the right hand side of (3), and $S(\tilde{z}|\mathcal{Z}_n)$ is the probability of estimating the PU traffic samples vector as $\tilde{z}$, when the actual PU traffic samples vector equals $\mathcal{Z}_n$. $S(\tilde{z}|\mathcal{Z}_n)$ can be written as

$$S(\tilde{z}|\mathcal{Z}_n) = P_f^{m_{0,n,\tilde{z}}}(1-P_f)^{m_{1,n,\tilde{z}}} P_m^{m_{2,n,\tilde{z}}}(1-P_m)^{m_{3,n,\tilde{z}}}, \quad (20)$$

where $m_{0,n,\tilde{z}}$, $m_{1,n,\tilde{z}}$, $m_{2,n,\tilde{z}}$, and $m_{3,n,\tilde{z}}$ are the numbers of false alarms, no false alarms, mis-detections, and no mis-detections, respectively, that yield the estimated PU traffic samples vector $\tilde{z}$ given the PU traffic samples vector $\mathcal{Z}_n$. The joint ML estimators of $u$ and $\lambda_f$, and $u$ and $\lambda_n$ can be modified to account for spectrum sensing errors. The modified ML estimators are calculated by solving for the values of $u$, $\lambda_f$, and $\lambda_n$ that maximize the modified likelihood function given in (19). The ML estimators as well as the corresponding mean squared estimation errors cannot be expressed in a simple closed-form and, hence, have to be calculated numerically as shown in Section V.

In the next section, the effect of sensing errors on the estimation error of $u$ is presented in a closed-form for the averaging estimator—an estimator that is commonly used in the literature.

## IV. Estimation of the Primary User Duty Cycle $u$ using Sample Averaging under Spectrum Sensing Errors

A common estimator for $u$ in the traffic estimation literature [3]–[7] is based on simply averaging the traffic samples. The estimator is easy to implement as it requires no a priori knowledge regarding $\lambda_f$ and $\lambda_n$. In this section we analyze the estimation error for the averaging estimator and we later compare it to that of the ML estimator in Section V. Furthermore, we derive closed-form expressions for the estimation error of the averaging estimator when spectrum sensing errors are considered. Moreover, the estimation error is analyzed for the general case where the traffic samples are not uniformly sampled. We first start by showing that spectrum sensing errors cause the estimator to be biased, then, we propose an unbiased estimator. Then, we present an expression for the mean estimation error as a function of the number of samples and the observation window length to serve as guidelines for traffic estimation in energy-constrained and delay-constrained systems, respectively. Finally, we show the dependence of the estimation error on spectrum sensing errors to provide intuition regarding the compromise between the time spent on spectrum sensing for each traffic sample and the estimation accuracy.

The biased averaging estimator can be expressed as $\tilde{u} = \frac{1}{N}\sum_{n=1}^{N}\tilde{z}_n$. The expected value of the estimator can be calculated as $E_{\tilde{z}}[\tilde{u}] = P_f(1-u) + u(1-P_m)$. Thus, the duty cycle can be calculated from $E_{\tilde{z}}[\tilde{u}]$ where $u = (E_{\tilde{z}}[\tilde{u}] - P_f)/(1 - P_f - P_m)$. Accordingly, the bias in the estimator can be eliminated where we propose the following unbiased estimator

$$\tilde{u}_a = \frac{1}{1-P_f-P_m}\left[-P_f + \frac{1}{N}\sum_{n=1}^{N}\tilde{z}_n\right]. \quad (21)$$

### A. The MSE in $\tilde{u}_a$

The MSE in $\tilde{u}_a$ for $N$ samples can be defined as $V_{\tilde{u}_a,N} = E_{\tilde{z}}[\tilde{u}_a^2] - u^2$. The expectation is calculated over all possible values of $\tilde{u}_a$ resulting from all $2^N$ permutations of the estimated traffic samples vector $\tilde{z}$. Define $\tilde{\mathcal{Z}}$ as a vector containing all $2^N$ permutations of $\tilde{z}$ with $\tilde{\mathcal{Z}}_n$, $n \in \{1,2,\cdots,2^N\}$, defined as the $n$th element of $\tilde{\mathcal{Z}}$. Define $\tilde{\mathcal{Z}}_{n,m}$, $m \in \{1,2,\cdots,N\}$, as the $m$th traffic sample of $\tilde{\mathcal{Z}}_n$. Thus, $V_{\tilde{u}_a,N}$ can be expressed as

$$V_{\tilde{u}_a,N} = \sum_{i=1}^{2^N} S_i^2 \Pr(\tilde{z} = \tilde{\mathcal{Z}}_i|\mathcal{T}) - u^2, \quad (22)$$

where $S_i = \frac{1}{1-P_f-P_m}\left[-P_f + \frac{1}{N}\sum_{j=1}^{N}\tilde{\mathcal{Z}}_{i,j}\right]$. We then have the following theorem.

*Theorem 1:* The MSE in $\tilde{u}_a$ is given as

$$V_{\tilde{u}_a,N} = \frac{2u(1-u)}{N^2}\sum_{i=1}^{N-1}\sum_{j=1}^{N-i}\prod_{k=j}^{i+j-1} e^{\frac{-T_k\lambda_f}{u}} + \frac{u(1-u)}{N}$$
$$+ \frac{uP_m(1-P_m) + (1-u)P_f(1-P_f)}{N(1-P_f-P_m)^2}. \quad (23)$$

*Proof:* See Appendix D. □

Considering the right hand side of (23), the leftmost term accounts for the estimation error caused by the sample correlation, while the rightmost term models the increase in the estimation error caused by spectrum sensing errors. Note that under perfect spectrum sensing, as $T_k$ tends to infinity, $V_{\tilde{u}_a,N}$ approaches $\frac{u(1-u)}{N}$, which is the MSE in estimating the duty cycle of an uncorrelated traffic sample sequence[2]. Finally, from (23) that the effect of the sensing error on the estimation error can be asymptotically eliminated by increasing $N$.

The work in [4]–[7] assumed a special case of the averaging estimator where uniform sampling is applied, that is, the inter-sample times are constant, $T_n = \frac{T}{N-1} = T_u$, $\forall T_n \in \mathcal{T}$. The MSE in $\tilde{u}_a$ under uniform sampling, denoted by $V_{\tilde{u}_{ua},N}$, can be derived by substituting $T_k = T_u$, $\forall T_k$, in (23).

For a fixed observation window length, as $N$ increases, the MSE error in estimating $u$ for uniform sampling approaches an asymptote, $V_{\tilde{u}_{ua},L}$, where

$$V_{\tilde{u}_{ua},L} = \lim_{N\to\infty} V_{\tilde{u}_{ua},N} = \frac{2u(1-u)}{\eta^2}\left(e^{-\eta} + \eta - 1\right), \quad (24)$$

where $\eta = \frac{T\lambda_f}{u}$. In order for $V_{\tilde{u}_{ua},L}$ to go to zero, $T$ has to be increased indefinitely. This implies that increasing the number of traffic samples while keeping the observation window length constant does not eliminate the estimation error caused by sample correlation.

## V. Numerical Results

The performance of the estimators of $u$, $\lambda_f$ and $\lambda_n$ is assessed in this section. Performance is quantified as the root mean squared (RMS) estimation error which is calculated as

---
[2]Note that $\frac{u(1-u)}{N}$ is the variance of a binomial distribution normalized by $N^2$ where the probability of success is set to $u$ [15, Ch. 4].

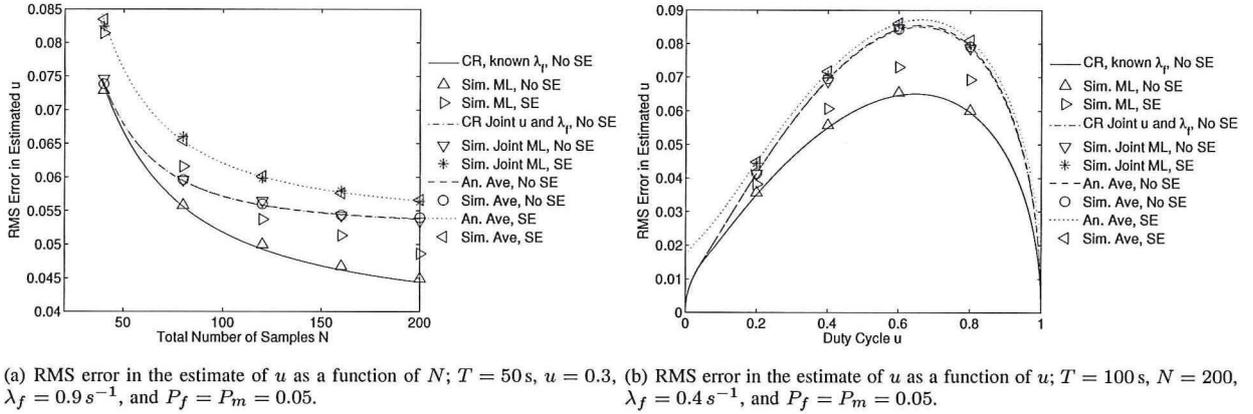

(a) RMS error in the estimate of $u$ as a function of $N$; $T = 50\,\text{s}$, $u = 0.3$, $\lambda_f = 0.9\,s^{-1}$, and $P_f = P_m = 0.05$.

(b) RMS error in the estimate of $u$ as a function of $u$; $T = 100\,\text{s}$, $N = 200$, $\lambda_f = 0.4\,s^{-1}$, and $P_f = P_m = 0.05$.

Fig. 1. RMS error in the estimate of $u$. Plots include analytical results for the CR bound with and without a priori knowledge of $\lambda_f$, the corresponding simulation-based ML estimation error results with and without spectrum sensing errors, and analytical and simulation results for the estimation error of the averaging estimator with and without spectrum sensing errors. SE: with sensing errors, No SE: without sensing errors, Ave: averaging estimator, An.: analysis, Sim.: simulation.

the square root of the mean squared estimation error. The variation of the RMS estimation error with the number of samples and the traffic parameters is plotted. Regarding the estimation of $u$, the RMS estimation error is compared for the averaging estimator, the blind joint ML estimator, and the ML estimator of $u$ assuming a priori knowledge of $\lambda_f$ that was presented in [3, Sec. III-C]. Considering the estimation of $\lambda_f$, the RMS estimation error for the blind joint ML estimator of $u$ and $\lambda_f$ is compared with that of the ML estimator proposed in [3, Sec. IV-A] that assumes perfect knowledge of $u$. Moreover, the square roots of the CR bounds on estimation accuracy are plotted in all figures as a reference. Note that the performance of the estimation of $\lambda_n$ is similar to that of $\lambda_f$, and thus is omitted to eliminate redundancy. Furthermore, the analysis presented in Section III-A and Section IV-A is verified using simulation data where typical traffic parameters were used following the results in [2], [7]. Finally, the impact of spectrum sensing errors on the RMS estimation error is demonstrated where analytical and simulation results are presented for the averaging estimator, while only simulation data is used for the ML estimators.

### A. Estimation of $u$

*1) The Variation of the RMS Error with the Number of Samples:* The variation of the RMS estimation error with the number of samples, $N$, is presented in Fig. 1(a) for $P_f = P_m = 0.05$. The RMS estimation error under perfect spectrum sensing is plotted in both figures as a reference. The total observation window length is 50 seconds and the assumed traffic parameters are $u = 0.3$ and $\lambda_f = 0.9\,s^{-1}$. The results show that, with zero spectrum sensing errors, the gap between the RMS estimation error, for all estimators, and the corresponding CR bounds is almost negligible. However, the gap increases with spectrum sensing errors. Moreover, simulation results show that for the considered spectrum sensing errors, the joint ML estimator and the averaging estimator yield almost the same error. Furthermore, the impact of the availability of a priori knowledge of $\lambda_f$ on the estimation error of $u$ is clear where the gap between the CR bounds on the estimation error with and without knowledge of $\lambda_f$ increases with $N$. Finally, the analytical expression for the estimation error for the averaging estimator presented in (23) is verified via Matlab simulations.

*2) The Variation of the RMS Error with the Duty Cycle $u$:* The RMS estimation error of $u$ as a function of the actual value of $u$ is plotted in Fig. 1(b) for $P_f = P_m = 0.05$. For this setup, $T = 100\,\text{s}$, $N = 200$ samples, and $\lambda_f = 0.4\,s^{-1}$. The results emphasize the fact that with zero spectrum sensing errors, the gap between the estimation error and the corresponding CR bounds for all estimators is almost negligible, where the gap increases with spectrum sensing errors. Besides, having a priori knowledge of $\lambda_f$ results in a significant decrease in the CR bound on the estimation error of $u$. However, the gain in incorporating a priori knowledge of $\lambda_f$ decreases as $u$ tends to 0 and 1. Finally, the estimation error expression for the averaging estimator presented in (23) is verified via Matlab simulations.

### B. Estimation of the Departure Rate $\lambda_f$

The relationship between the RMS estimation error in $\lambda_f$ with the number of samples is plotted in Fig. 2(a), while the variation of the estimation error with $\lambda_f$ is shown in Fig. 2(b). The results are presented for the estimation of $\lambda_f$ with and without having a priori knowledge of $u$. The figures show that the two ML estimators yield RMS estimation errors that are within a narrow gap from their corresponding CR bounds, where the gap increases with spectrum sensing errors. Moreover, the estimation error increases with $\lambda_f$. Finally, the results emphasize the fact that having a priori knowledge of $u$ results in a notable decrease in the estimation error of $\lambda_f$.

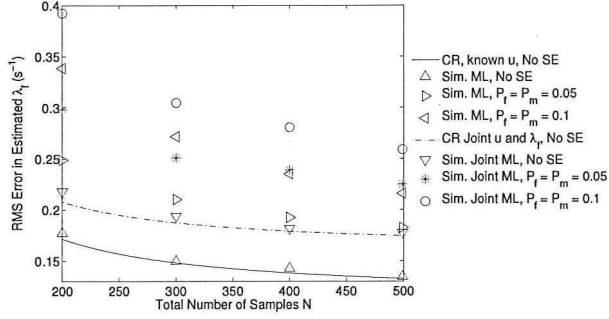

(a) RMS error in the estimate of $\lambda_f$ as a function of $N$; $u = 0.3$, $\lambda_f = 0.9\,s^{-1}$, $T = 50\,s$.

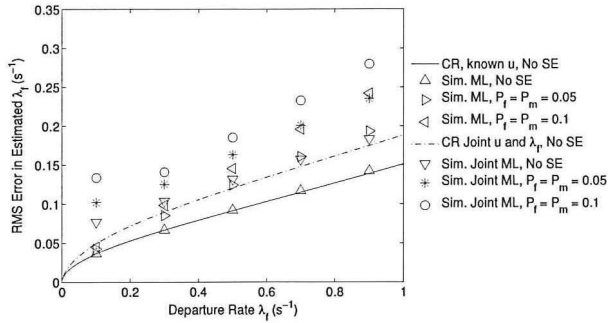

(b) RMS error in the estimate of $\lambda_f$ as a function of $\lambda_f$; $u = 0.6$, $T = 100\,s$, $N = 300$.

Fig. 2. RMS error in the estimate of $\lambda_f$. Plots include analytical results for the CR bound with and without a priori knowledge of $u$, and the corresponding simulation-based ML estimation error results without sensing errors, and with sensing errors of $P_f = P_m = 0.05$ and $P_f = P_m = 0.1$. SE: with sensing errors, No SE: without sensing errors, An.: analysis, Sim.: simulation.

## VI. CONCLUSIONS

In this paper, we developed a mathematical framework for quantifying the estimation accuracy of PU traffic parameters with exponentially distributed off- and on-times. We derived the Cramér-Rao (CR) bounds on the traffic estimates when uniform sampling is used. The CR bounds consider the blind joint estimation of all traffic parameters. We showed that, due to the sample correlation, the CR bounds approach a lower bound as the number of samples increase, where the bounds can only be reduced by increasing the total observation window length. We also proved that the effect of having a priori knowledge of the mean PU arrival or departure rates on the asymptotic CR bound on estimating $u$ is equivalent to doubling the total observation window length. Besides, we presented expressions for the joint maximum-likelihood estimators of $u$ and the mean departure rate, and showed via simulations that they approach the CR bounds asymptotically. Moreover, we formulated the modified likelihood function for the joint estimation of traffic parameters when spectrum sensing errors are considered. Furthermore, we analyzed the effect of spectrum sensing errors on the estimation of $u$ when sample averaging is applied. We proved that as the number of samples increases, the effect of spectrum sensing errors on the estimation error diminishes, yet, the estimation error is lower bounded as a result of the traffic samples correlation. Finally, we showed via simulations that the gap between the estimation error for the averaging estimator with no spectrum sensing errors and the corresponding CR bound is negligible.

## APPENDIX A

*Proof:* First, let $\phi_1 = \frac{-\partial^2}{\partial u^2} \log L(\boldsymbol{z}|\boldsymbol{\theta})$, then, $\phi_1$ can be expressed as $\frac{u^2 - 2z_1 u + z_1}{u^2(1-u)^2} + n_0\Omega_0 + n_1\Omega_1 + n_2\Omega_2 + n_3\Omega_3$, where

$$\Omega_0 = \frac{u\Pr_{01}^2 - 2uT_c\lambda_f\Gamma_c\Pr_{01} - T_p^2\lambda_f^2\Gamma_c(1-u)}{u^3\Pr_{00}^2}, \quad (A.25)$$

$$\Omega_1 = \frac{\Pr_{01}^2 + T_c\lambda_f\Gamma_c(T_c\lambda_f - 2\Pr_{01})}{u^2\Pr_{01}^2}, \quad (A.26)$$

$$\Omega_2 = \frac{u^2\Pr_{01}^2 + T_c\lambda_f\Gamma_c(1-u)^2(T_c\lambda_f - 2\Pr_{01})}{u^2(1-u)^2\Pr_{01}^2}, \quad (A.27)$$

and

$$\Omega_3 = \frac{u\Pr_{01}^2 + 2T_c\lambda_f\Gamma_c(2\Pr_{11} - u\Pr_{01} - \Gamma_c)}{u^3\Pr_{11}^2}$$
$$- \frac{\Gamma_c T_c^2\lambda_f^2(1-u)}{u^3\Pr_{11}^2}, \quad (A.28)$$

where the argument of $\Pr_{xy}(T_c|u, \lambda_f)$, $x, y \in \{0, 1\}$, is dropped for brevity. The expression presented in (9) can be derived by mathematical induction as follows. Define $\mathcal{Z} = [\mathcal{Z}_1, \mathcal{Z}_2, \cdots, \mathcal{Z}_{2^N}]$ as a vector containing all $2^N$ permutations of $\boldsymbol{z}$. Then, define $\Phi_1$ and $P$ as vectors containing the corresponding values of $\phi_1$ and $\Pr(\boldsymbol{z} = \mathcal{Z}_n|\mathcal{T})$ (the probability of observing the traffic sample vector $\mathcal{Z}_n$ given $\mathcal{T}$), respectively. Finally, define $\mathcal{Z}_n$, $\Phi_{1,n}$, and $P_n$ as the $n$th elements of $\mathcal{Z}$, $\Phi_1$, and $P$, respectively. Hence, $E_{\boldsymbol{z}}[\phi_1] = \sum_{i=1}^{2^N} \Phi_{1,i} P_i$. For the base case with $N = 2$, $\mathcal{Z} = [00, 01, 10, 11]$, $\Phi_1 = [(1-u)^{-2} + \Omega_0, (1-u)^{-2} + \Omega_1, u^{-2} + \Omega_2, u^{-2} + \Omega_3]$, and $P = [(1-u)\Pr_{00}, (1-u)\Pr_{01}, u\Pr_{10}, u\Pr_{11}]$. Thus,

$$E_{\boldsymbol{z}}[\phi_1] = \frac{\lambda_f T_c(1-u)(1+\Gamma_c) + (4u^2 - 2u)(1-\Gamma_c)}{u^2\Pr_{01}\Pr_{00}\Pr_{11}}$$
$$\times \Gamma_c^2\lambda_f T_c - \frac{2\Pr_{10} u(2\Gamma_c - 1) + \Gamma_c(\Gamma_c - 2)}{u(1-u)\Pr_{00}\Pr_{11}}, \quad (A.29)$$

which equals (9) for $N = 2$, hence, confirms the base case.

Assuming that (9) is true for $N$, proving that the expression holds for $N + 1$ is sufficient for proving (9). For notation simplicity, to differentiate between the cases with $N$ and $N + 1$ samples, we add the number of samples as a superscript, denoted by $(N)$ and $(N+1)$, respectively, for $E_{\boldsymbol{z}}[\phi_1]$, $\mathcal{Z}$, $\mathcal{Z}_n$, $P_n$, $n_0$, $n_1$, $n_2$, and $n_3$. Accordingly,

$$E_{\boldsymbol{z}}^{(N+1)}[\phi_1] = \sum_{i=1}^{2^{N+1}} P_i^{(N+1)} \Big( \frac{u^2 - 2z_1 u + z_1}{u^2(1-u)^2} + n_0^{(N+1)}\Omega_0$$
$$+ n_1^{(N+1)}\Omega_1 + n_2^{(N+1)}\Omega_2 + n_3^{(N+1)}\Omega_3 \Big). \quad (A.30)$$

The vector of all traffic sample sequences of length $N$, $\mathcal{Z}^{(N)}$, can be split to $\mathcal{Z}^{(N,0)}$ and $\mathcal{Z}^{(N,1)}$, which represent the traffic sample sequences of length $N$ ending with 0 and 1, respectively. Thus, $E_{\mathbf{z}}^{(N+1)}[\phi_1]$ can be expressed as

$$E_{\mathbf{z}}^{(N+1)}[\phi_1] = \sum_{\mathcal{Z}_i^{(N)} \in \mathcal{Z}^{(N,0)}} P_i^{(N)} \left[ \Pr_{00}(\Psi_1 + \Omega_0) + \Pr_{01}(\Psi_1 + \Omega_1) \right]$$
$$+ \sum_{\mathcal{Z}_i^{(N)} \in \mathcal{Z}^{(N,1)}} P_i^{(N)} \left[ \Pr_{10}(\Psi_1 + \Omega_2) + \Pr_{11}(\Psi_1 + \Omega_3) \right], \quad \text{(A.31)}$$

where $\Psi_1 = \frac{u^2 - 2z_1 u + z_1}{u^2(1-u)^2} + n_0^{(N)}\Omega_0 + n_1^{(N)}\Omega_1 + n_2^{(N)}\Omega_2 + n_3^{(N)}\Omega_3$. Thus, $E_{\mathbf{z}}^{(N+1)}[\phi_1] = E_{\mathbf{z}}^{(N)}[\phi_1] + (1-u)(\Pr_{00}\Omega_0 + \Pr_{01}\Omega_1) + u(\Pr_{10}\Omega_2 + \Pr_{11}\Omega_3)$. It follows that

$$E_{\mathbf{z}}^{(N+1)}[\phi_1] = \frac{\lambda_f T_c(1-u)(1+\Gamma_c) + (4u^2 - 2u)(1-\Gamma_c)}{u^2 \Pr_{01} \Pr_{00} \Pr_{11}}$$
$$\times \Gamma_c^2 \lambda_f T_c N - \frac{u \Pr_{10}((3N+1)\Gamma_c - N - 1)}{u(1-u)\Pr_{00}\Pr_{11}}$$
$$+ \frac{\Gamma_c(N\Gamma_c - N - 1)}{u(1-u)\Pr_{00}\Pr_{11}}, \quad \text{(A.32)}$$

which corresponds to (9) with the number of samples set to $N+1$, hence, the derivation is concluded. $\square$

## Appendix B

*Proof:* The derivation for (10) is similar to that of (9). First, let $\phi_2$ denote $\frac{-\partial^2}{\partial u \partial \lambda_f} \log L(z|\theta)$, then, $\phi_2$ can be expressed as $n_0 \Upsilon_1 + (n_1 + n_2)\Upsilon_2 + n_3 \Upsilon_3$, where

$$\Upsilon_1 = \frac{\Gamma_c T_p \left(u^2(1-\Gamma_c) + T_c \lambda_f (1-u)\right)}{u^2 \Pr_{00}^2}, \quad \text{(B.33)}$$

$$\Upsilon_2 = \frac{\Gamma_c T_c(u(1-\Gamma_c) - T_c \lambda_f)}{u \Pr_{01}^2}, \quad \text{(B.34)}$$

and

$$\Upsilon_3 = \frac{T_c \Gamma_c \left(u(u-2) + T_c \lambda_f(1-u) - \Gamma_c(1-u)^2\right)}{u^2 \Pr_{11}^2}. \quad \text{(B.35)}$$

The expression presented in (10) can be derived by mathematical induction as follows where $\mathcal{Z}$, $\mathcal{Z}_n$, $P$, and $P_n$ are as defined in Appendix A. Moreover, define $\Phi_2$ as a vector where $\Phi_{2,n}$, the $n$th element of $\Phi_2$, equals $\phi_2$ evaluated for $\mathcal{Z}_n$. Hence, $E_{\mathbf{z}}[\phi_2] = \sum_{i=1}^{2^N} \Phi_{2,i} P_i$. For the base case with $N = 2$, $\mathcal{Z}$ and $P$ are as defined in Appendix A. Furthermore, $\Phi_2 = [\Upsilon_1, \Upsilon_2, \Upsilon_2, \Upsilon_3]$. Thus, $E_{\mathbf{z}}[\phi_2]$ can be calculated as

$$E_{\mathbf{z}}[\phi_2] = \frac{T_c \Gamma_c^2 (\Pr_{01}(1 - 2u) - T_c \lambda_f(1-u)(1+\Gamma_c))}{u \Pr_{01} \Pr_{11} \Pr_{00}}, \quad \text{(B.36)}$$

which equals (10) for $N = 2$, hence, the base case is confirmed.

Assuming that (10) is true for $N$, proving that the expression holds for $N+1$ is sufficient for proving (10). As in Section A, the superscripts $(N)$ and $(N+1)$ are used to differentiate between cases with $N$ and $N+1$ samples, respectively. Thus,

$$E_{\mathbf{z}}^{(N+1)}[\phi_2] = \sum_{i=1}^{2^{N+1}} P_i^{(N+1)} \left[ n_0^{(N+1)} \Upsilon_1 + \left(n_1^{(N+1)} + n_2^{(N+1)}\right)\Upsilon_2 + n_3^{(N+1)} \Upsilon_3 \right]$$
$$= \sum_{\mathcal{Z}_i^{(N)} \in \mathcal{Z}^{(N,0)}} P_i^{(N)} \left[ \Pr_{00}(\Psi_2 + \Upsilon_1) + \Pr_{01}(\Psi_2 + \Upsilon_2) \right]$$
$$+ \sum_{\mathcal{Z}_i^{(N)} \in \mathcal{Z}^{(N,1)}} P_i^{(N)} \left[ \Pr_{10}(\Psi_2 + \Upsilon_2) + \Pr_{11}(\Psi_2 + \Upsilon_3) \right],$$
(B.37)

where $\Psi_2 = n_0^{(N)} \Upsilon_1 + (n_1^{(N)} + n_2^{(N)})\Upsilon_2 + n_3^{(N)}\Upsilon_3$. It follows that $E_{\mathbf{z}}^{(N+1)}[\phi_2] = E_{\mathbf{z}}^{(N)}[\phi_2] + (1-u)(\Pr_{00}\Upsilon_1 + \Pr_{01}\Upsilon_2) + u(\Pr_{10}\Upsilon_2 + \Pr_{11}\Upsilon_3)$. Accordingly

$$E_{\mathbf{z}}^{(N+1)}[\phi_2] = \frac{T_c \lambda_f(1-u)(1+\Gamma_c) + \Pr_{01}(2u-1)}{u \Pr_{01} \Pr_{00} \Pr_{11}}$$
$$\times -N\Gamma_c^2 T_c, \quad \text{(B.38)}$$

which corresponds to (10) with the number of samples set to $N+1$, hence, the derivation is concluded. $\square$

## Appendix C

*Proof:* The derivation of expression (11) follows that of (9). Denote $\frac{-\partial^2}{\partial \lambda_f^2} \log L(z|\theta)$ by $\phi_3$, hence, $\phi_3$ can be expressed as $n_0 \Lambda_1 + (n_1 + n_2)\Lambda_2 + n_3 \Lambda_3$, where $\Lambda_1 = \left(-(1-u)\Gamma_c T_c^2\right)/\left(u \Pr_{00}^2\right)$, $\Lambda_2 = \left(\Gamma_c T_c^2\right)/\left(\Pr_{01}^2\right)$, and $\Lambda_3 = \left(-(1-u)\Gamma_c T_c^2\right)/\left(u \Pr_{11}^2\right)$. The expression presented in (11) can be shown by mathematical induction as follows where $\mathcal{Z}$, $\mathcal{Z}_n$, $P$, and $P_n$ are as defined in Appendix A. Moreover, define $\Phi_3$ as a vector where $\Phi_{3,n}$, the $n$th element of $\Phi_3$, equals $\phi_3$ evaluated for $\mathcal{Z}_n$. Hence, $E_{\mathbf{z}}[\phi_3] = \sum_{i=1}^{2^N} \Phi_{3,i} P_i$. For the base case with $N = 2$, $\mathcal{Z}$ and $P$ are as defined in Appendix A. Moreover, $\Phi_3 = [\Lambda_1, \Lambda_2, \Lambda_2, \Lambda_3]$. Thus, $E_{\mathbf{z}}[\phi_3]$ can be calculated as $\left(\Gamma_c^2 T_c^2 (1-u)(1+\Gamma_c)\right)/(\Pr_{01} \Pr_{00} \Pr_{11})$, which equals (11) for $N = 2$, hence, it concludes the base case.

The expression presented in (11) can be derived by showing that (11) holds for $N+1$ under the assumption that (11) holds for $N$. As in Section A, superscripts $(N)$ and $(N+1)$ are used to differentiate between cases with $N$ and $N+1$ samples, respectively. Thus,

$$E_{\mathbf{z}}^{(N+1)}[\phi_3] = \sum_{i=1}^{2^{N+1}} P_i^{(N+1)} \left[ n_0^{(N+1)} \Lambda_1 + (n_1^{(N+1)} + n_2^{(N+1)})\Lambda_2 + n_3^{(N+1)} \Lambda_3 \right]$$
$$= \sum_{\mathcal{Z}_i^{(N)} \in \mathcal{Z}^{(N,0)}} P_i^{(N)} \left[ \Pr_{00}(\Psi_3 + \Lambda_1) + \Pr_{01}(\Psi_3 + \Lambda_2) \right]$$
$$+ \sum_{\mathcal{Z}_i^{(N)} \in \mathcal{Z}^{(N,1)}} P_i^{(N)} \left[ \Pr_{10}(\Psi_3 + \Lambda_2) + \Pr_{11}(\Psi_3 + \Lambda_3) \right],$$
(C.39)

where $\Psi_3 = n_0^{(N)}\Lambda_1 + (n_1^{(N)} + n_2^{(N)})\Lambda_2 + n_3^{(N)}\Lambda_3$. It follows that $E_{\tilde{z}}^{(N+1)}[\phi_3] = E_{\tilde{z}}^{(N)}[\phi_3] + (1-u)(\Pr_{00}\Lambda_1 + \Pr_{01}\Lambda_2) + u(\Pr_{10}\Lambda_2 + \Pr_{11}\Lambda_3)$. Accordingly, $E_{\tilde{z}}^{(N+1)}[\phi_3]$ can be expressed as $\left(N\Gamma_c^2 T_c^2(1-u)(1+\Gamma_c)\right)/(\Pr_{01}\Pr_{00}\Pr_{11})$, which corresponds to (11) with the number of samples set to $N+1$, hence, proving (11) by mathematical induction. □

## APPENDIX D

*Proof:* The expression presented in (23) can be proved by mathematical induction as in Appendix A. Define $\tilde{P}_n = \Pr(\tilde{z} = \tilde{\mathcal{Z}}_i | \mathcal{T})$, i.e., the probability of observing the estimated PU traffic samples vector $\tilde{\mathcal{Z}}_i$ given $\mathcal{T}$. Moreover, define $\tilde{P}$ as a vector containing all values of $\tilde{P}_n$ such that $\tilde{P}_n$ is the $n$th element of $\tilde{P}$. Furthermore, define $\mathcal{S}$ as a vector where $S_n$, defined in Section IV-A, is the $n$th element of $\mathcal{S}$. Hence, $V_{\tilde{u}_a,N} = \sum_{i=1}^{2^N} S_i^2 \tilde{P}_i - u^2$.

For the base case with $N=2$, $\tilde{\mathcal{Z}} = [00, 01, 10, 11]$, where $\tilde{\mathcal{Z}}$ is as defined in Section IV-A, and $\mathcal{S} = -1/(1-P_f-P_m)[-P_f, -P_f+1/2, -P_f+1/2, -P_f+1]$. Besides, $\tilde{P} = [(1-u)\bar{P}_f[\Pr_{00}\bar{P}_f + \Pr_{01}\bar{P}_m] + uP_m[\Pr_{10}\bar{P}_f + \Pr_{11}\bar{P}_m], (1-u)\bar{P}_f[\Pr_{00}P_f + \Pr_{01}\bar{P}_m] + uP_m[\Pr_{10}P_f + \Pr_{11}\bar{P}_m], (1-u)P_f[\Pr_{00}\bar{P}_f + \Pr_{01}P_m] + uP_m[\Pr_{10}\bar{P}_f + \Pr_{11}P_m], (1-u)P_f[\Pr_{00}P_f + \Pr_{01}\bar{P}_m] + u\bar{P}_m[\Pr_{10}P_f + \Pr_{11}\bar{P}_m]]$, where $\bar{P}_f = 1 - P_f$ and $\bar{P}_m = 1 - P_m$. Accordingly,

$$V_{\tilde{u}_a,2} = \frac{u(1-u)\Gamma_c}{2} + \frac{u(1-u)}{2} + \frac{uP_m(1-P_m) + (1-u)P_f(1-P_f)}{2(1-P_f-P_m)^2}, \quad \text{(D.40)}$$

which equals (23) for $N=2$, hence, proves the base case.

Showing that (23) holds for $N+1$ while assuming that it is true for $N$ is sufficient for proving (23). Subscripts $(N)$ and $(N+1)$ are added to $P_n$ and $S_n$ to differentiate between cases with $N$ and $N+1$ samples, respectively. For $N+1$ samples, $V_{\tilde{u}_a,N+1} = \sum_{i=1}^{2^{N+1}} S_{i,(N+1)}^2 P_{i,(N+1)} - u^2$ and can be expressed as

$$V_{\tilde{u}_a,N+1} = \frac{N^2 V_{\tilde{u}_a,N}}{(N+1)^2} + \Theta_1 + \Theta_2 - \frac{(2N+1)u^2}{(N+1)^2}, \quad \text{(D.41)}$$

where

$$\Theta_1 = \sum_{i=1}^{2^{N+1}} P_{i,(N+1)} \left(\sum_{n=1}^{N} \tilde{\mathcal{Z}}_{i,n} - NP_f\right)(\tilde{\mathcal{Z}}_{i,N+1} - P_f) \times \frac{2}{(N+1)^2(1-P_f-P_m)^2}, \quad \text{(D.42)}$$

and

$$\Theta_2 = \frac{1}{(N+1)^2(1-P_f-P_m)^2} \times \sum_{i=1}^{2^{N+1}} P_{i,(N+1)}(\tilde{\mathcal{Z}}_{i,N+1} - P_f)^2. \quad \text{(D.43)}$$

Note that $\tilde{\mathcal{Z}}_{i,N+1}$ in $\Theta_1$ and $\Theta_2$ denotes the estimated traffic sample number $N+1$ in the traffic sample vector $\tilde{\mathcal{Z}}_i$. $\Theta_1$ is a recursive expression that can be simplified to

$$\Theta_1 = \frac{2Nu^2}{(N+1)^2} + \frac{2u(1-u)}{(N+1)^2}\sum_{j=1}^{N}\prod_{k=j}^{N}e^{\frac{-T_k\lambda_f}{u}}, \quad \text{(D.44)}$$

where the proof is omitted for brevity. Moreover, $\Theta_2$ can be simplified to

$$\Theta_2 = \frac{P_f(1-P_f) + u(1-2P_f)(1-P_f-P_m)}{(N+1)^2(1-P_f-P_m)^2}. \quad \text{(D.45)}$$

Finally, substituting (23) for $V_{\tilde{u}_a,N}$ in (D.41) and using the simplified expressions for $\Theta_1$ and $\Theta_2$, we obtain

$$V_{\tilde{u}_a,N+1} = \frac{2u(1-u)}{(N+1)^2}\sum_{i=1}^{N}\sum_{j=1}^{N+1-i}\prod_{k=j}^{i+j-1}e^{\frac{-T_k\lambda_f}{u}} + \frac{u(1-u)}{N+1} + \frac{uP_m(1-P_m) + (1-u)P_f(1-P_f)}{(N+1)(1-P_f-P_m)^2}. \quad \text{(D.46)}$$

This corresponds to (23) with the number of samples set to $N+1$, hence, proves (23) by mathematical induction. □